\documentclass[%
 aip,
 amsmath,amssymb,
 reprint,%
]{revtex4-1}

\usepackage{graphicx}
\usepackage{epsfig}

\begin{document}
\title{Circuit simulation of readout process toward large-scale superconducting quantum circuits}

\author{Tetsufumi Tanamoto}

\affiliation{Department of Data Science, Teikyo University,
2-11-1 Kaga, Itabashi-ku, Tokyo, 173-8605, Japan}

\author{Hiroshi Fuketa}
\address{
%Semiconductor Frontier Research Center (SFRC), 
National Institute of Advanced Industrial Science and Technology (AIST), Tsukuba
Ibaraki 305-8568, Japan}
\author{Toyofumi Ishikawa}
\address{
%Research Center for Emerging Computing Technologies (RCECT), 
National Institute of Advanced Industrial Science and Technology (AIST), Tsukuba
Ibaraki 305-8568, Japan}
\author{Shiro Kawabata}
\address{
Faculty of Computer and Information Sciences,
Hosei University,
3-7-2 Kajino, Koganei, Tokyo 184-8584, Japan}

\begin{abstract}
The rapid scaling of superconducting quantum computers has highlighted the impact of device-level variability on overall circuit fidelity. 
In particular, fabrication-induced fluctuations in device parameters such as capacitance and Josephson critical current pose significant challenges to large-scale integration. 
We propose a simulation methodology for estimating qubit fidelity based on classical circuit simulation, 
using a conventional Simulation Program with Integrated Circuit Emphasis (SPICE) simulator.  
This approach enables the evaluation of the performance of superconducting quantum circuits with 10000 qubits
on standard laptop computers. 
The proposed method provides an accessible tool for the early stage assessment of large-scale superconducting quantum circuit performance.
\end{abstract}
\maketitle

%\section{Introduction}
The ongoing advancements in superconducting quantum computers have steadily increased the number of operational qubits, 
with current systems exceeding 100 ~\cite{Fujii,Mohseni,Brennan}.
Furthermore, quantum error correction beyond the break-even point has recently been achieved and experimentally demonstrated~\cite{Google,Google1,Lacroix,Yale,Ni}. 
Despite these advances, practical quantum computing applications require fault-tolerant quantum computers comprising millions of physical qubits~\cite{Gidney1}.
The qubit number scale and circuit complexity necessitate new design methodologies and scalable simulation techniques.

In this context, simulations play a pivotal role in the development of integrated quantum systems. 
Circuit-level simulations based on conventional complementary metal--oxide--semiconductor (CMOS) technology are essential, 
as superconducting quantum computers are driven by classical control electronics, 
including cryogenic CMOS-based circuits~\cite{Chakraborty,Sadhu} and quantum error correction decoders~\cite{Battistel}.

Two complementary types of simulations are required to design large-scale quantum circuits: rapid, coarse-grained simulations for early stage analysis and fine-grained simulations for optimizing detailed circuit parameters. 
Simulations that incorporate quantum effects often require substantial computational resources\cite{WRSPICE,PSCAN,JSIM}. 
In contrast, simplified classical simulations must be fast and scalable. 
We focus on simplified simulations in which qubits are represented as LCR circuits.
In our previous study~\cite{TanaAPE}, we introduced a Simulation Program with Integrated Circuit Emphasis (SPICE)-based method to estimate the qubit relaxation time $T_1$ in superconducting circuits. 
However, the circuit performance is more directly characterized by fidelity, particularly in the context of a readout.

In this study, we extended our simulation framework to enable fidelity estimation using classical SPICE simulations. 
We targeted transmon qubits~\cite{Koch} 
in the dispersive readout region~\cite{Wallraff,Blais,Krantz,Jeffery,Goppl,Houck,Riste,Nakamura,Materise}.
Our objective was to evaluate the superconducting-quantum-circuit performance using classical techniques wherever feasible. 
Simulations involving 1,000 qubits can be completed within minutes on a standard laptop computer (13th Gen Intel Core i7-1355U, 1.70 GHz), whereas those involving 10,000 qubits require approximately 2 h. Our analysis focuses on the influence of parameter variations, such as capacitance and resistance, on readout fidelity. 
We deliberately maintained a simple circuit architecture to support a large-scale evaluation.

Randomized benchmarking (RB) and cross-entropy benchmarking (XEB) are widely used 
to assess the quality of quantum computer hardware~\cite{Knill,Ryan}.
Whereas RB measures the average error rate of quantum gates, XEB quantifies the fidelity of quantum circuits. 
However, RB is not directly applicable within the scope of our classical simulations, 
as RB involves arbitrary single-qubit rotations and two qubit gates, 
features that cannot be represented within classical frameworks. 
Thus, in this study, we assume that quantum mechanical simulations 
were applied to evaluate the intrinsic coherence parameters $T_1$ and $T_2$.
We then replace the conventional XEB estimation process with circuit-level simulations based on classical modeling.

%%%%%%%%%%%%%%%%%%%%%%%%%%%%%%%%%%%%  Fig.1
\begin{figure}
\centering
\includegraphics[width=7cm]{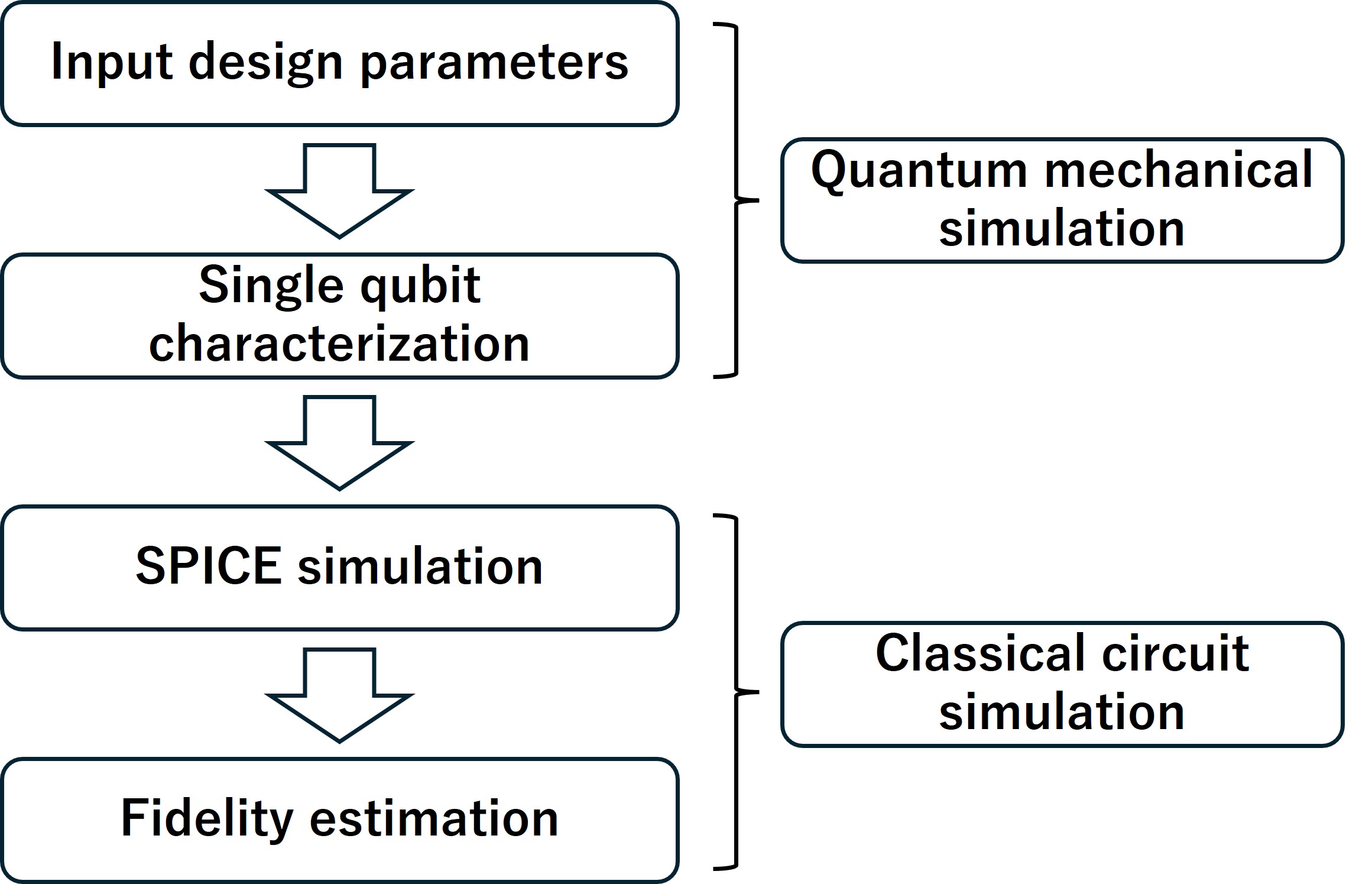}
\caption{
Flow to estimate the fidelity of the system.
Quantum mechanical simulation is performed for one and two qubits 
to extract qubit parameters (relaxation rates) for the following circuit simulations.
The fidelity is estimated based on the circuit simulations using SPICE. 
The quantum mechanical simulation part 
can be omitted when the qubit parameters are appropriately prepared.
}
\label{flow}
\end{figure}
%%%%%%%%%%%%%%%%%%%%%%%%%%%%%%%%%%%%
%\section{Flow}

Figure 1 illustrates the flowchart for estimating the fidelity of superconducting circuits. 
Decoherence is characterized by longitudinal and transverse relaxation times, 
denoted as \(T_1\) and \(T_2\), which correspond to the longitudinal and transverse relaxation rates, \(\Gamma_1\) and \(\Gamma_2\). 
We applied the Bloch--Redfield theory to describe the decoherence process~\cite{Krantz}, assuming a weak coupling between the qubit and its environment. 
This process was executed using a numerical approach, which is discussed later.

%%\subsection{Estimation of Fidelity}
The fidelity is calculated using the formulation derived by Abad {\it et al},
 as described in\cite{Abad}
\begin{equation}
F=1-\frac{N2^N}{2(2^N+1)}\tau_{\rm op}^{(0)} [\Gamma_1+\Gamma_2].
\label{fidelity}
\end{equation}
where $\tau_{\rm op}^{(0)}$ is the operating time.
$\Gamma_1$ and $\Gamma_2$ represent the relaxation and dephasing times, respectively.
This equation was used to estimate the fidelity of the output. 
Here, we consider \(\tau_{\rm op}^{(0)}\) as an adjustment parameter, 
which should be determined experimentally in the future. 
In addition, we assume that \(\Gamma_2 \approx 2\Gamma_1\), 
which is included in the operation time $\tau_{\rm op}$, 
such that \([\Gamma_1+\Gamma_2] \tau_{\rm op}^{(0)} \approx \Gamma_1 \tau_{\rm op}\).
The prefactor \(\frac{N2^N}{2(2^N+1)}\) in Equation (\ref{fidelity}) significantly influences fidelity values. 
For instance, when \(N=1\), the prefactor is \(\frac{1}{3}\). 
However, as \(N\) increases, the prefactor approaches \(\frac{N}{2}\), 
leading to fidelity degradation.

%%%%%%%%%%%%%%%%%%%%%%%%%%%%%%%%%%%%  Fig.2
\begin{figure}
\centering
\includegraphics[width=8.8cm]{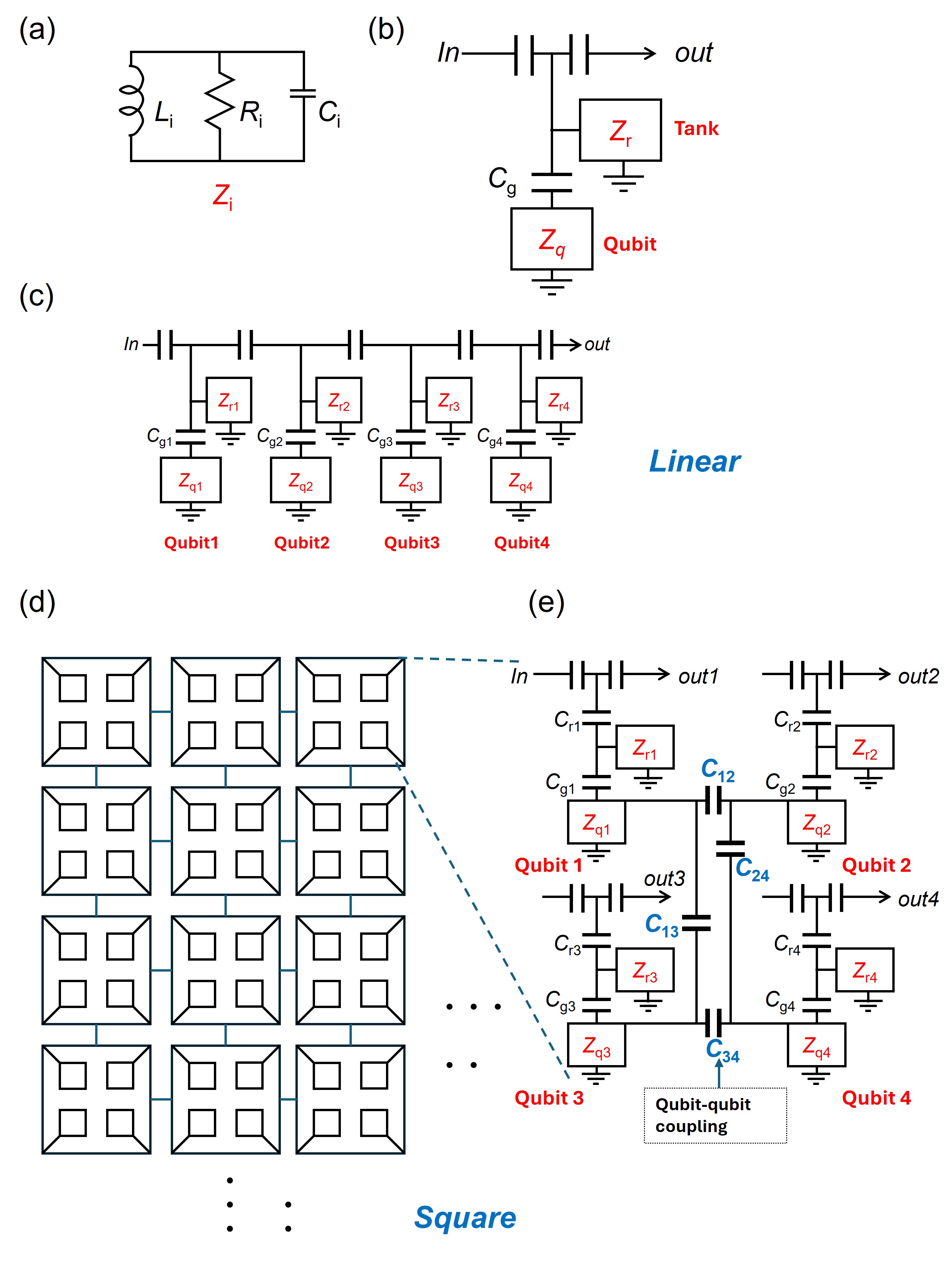}
\caption{
Classical circuit models for the transmon system.
(a) Basic unit of the transmon qubit and tank circuit.
(b) A single circuit which describes the qubit with the tank circuit.
(c) Measurement circuit describing linear qubit array.
The qubits have the frequencies $\omega_{q1}/2\pi=6$,
$\omega_{q2}/2\pi=6.2$, $\omega_{q3}/2\pi=6.4$, and $\omega_{q4}/2\pi=6.6$ GHz.
The tank circuits have the frequencies 
$\omega_{r1}/2\pi=8$ GHz,
$\omega_{r2}/2\pi=8.2$ GHz, $\omega_{r3}/2\pi=8.4$ Hz, and $\omega_{r4}/2\pi=8.6$ GHz.
(d) Measurement circuit describing the square arrangement of the qubits.
(e) Unit of the square arrangement of qubits.
The qubits have the frequency $\omega_{q1}/2\pi=6$ GHz, 
whereas the tank circuits have the frequencies of $\omega_{r1}/2\pi=8$ GHz.
In all cases, $T_1^{(q)}=1~\mu$s, $C_{q}=$30 fF, and $C_{r}=$20 fF.
The coupling capacitance between the qubit and transmission line is given by $C_{g}=5$~fF.
The coupling between qubits ($C_{12}$,$C_{23}$,$C_{34}$, and $C_{41}$) is given by 0.1 fF capacitances. 
}
\label{qubit_array}
\end{figure}
%%%%%%%%%%%%%%%%%%%%%%%%%%%%%%%%%%%%

%\section{Model}
Figure~\ref{qubit_array} illustrates the qubit array considered in this study. 
We consider the typical setup of a cavity QED in which 
qubit was capacitively coupled to a transmission line~\cite{Wallraff}.
The qubit and tank circuits are described by LCR circuits, as shown in Fig.2(a).
Figure 2(b) shows the basic measurement unit, which comprises a qubit ($Z_q$) and a tank circuit ($Z_r$).
Several circuit architectures have been proposed~\cite{IBMlayout,Heinsoo}; 
however, two types of simple qubit architectures were examined. 
In Fig.~2(c), four qubits are measured using a single transmission line ("$linear$ $structure$"). 
In contrast, Fig.~2(d) illustrates a structure where each qubit is accessed independently ("$square$ $structure$").
As shown in Fig.~2(d), the AC source is connected to one of the four qubits. 
This setup was necessary for the measurement phase of the surface code, where one of the qubits was measured~\cite{Fowler}. 

The number of parameters in each circuit unit should be small to calculate the performance of larger circuits.
We consider four input parameters: qubit frequency $\omega_q$, qubit relaxation time $T_1^{(q)}$, and 
and the capacitance in qubit $C_q$.
$T_1^{(q)}$ is calculated from the quantum mechanical simulation process shown in Fig.~1.
Here, we take $T_1^{(q)}$ as a given parameter for simplicity.
We consider the measurement process in the dispersive readout~\cite{Blais,Krantz}.
The photons in the transmission line interact with the qubit in terms of the energy difference $\omega_q$ between the 
the ground and first excited states~\cite{Blais}. 
The dissipation is assumed to be described by the resistance $R_q$ in the LCR circuits
with the relationship $T_1^{(q)}=C_qR_q$~\cite{Jeffery,Houck} given by $R_q=T_1^{(q)}/C_q$. 
The capacitances $C_q$ are of the order of 10 fF, and the frequencies $\omega_q/2\pi$ are of the order of GHz.
Thus, $C_q\omega_q/2\pi \sim 10~{\rm fF}\times {\rm GHz} \sim 10^{-5}$ $\Omega^{-1}$.
Assuming $T_1^{(q)}=1~\mu$s, 
the resistance $R_q$ can be expressed as: 
$1/R_q\sim C_q/T_1\sim 10~{\rm fF} /1~\mu {\rm s}
=10^{-8}$ $\Omega^{-1}$, and $1/R_q \ll C_q\omega_q$ are obtained. 
The inductance $L_q$ ($L_r$) of the qubit circuit (tank circuit) is given 
by $L_q =1/(C_q \omega_q^2)$ ($L_r =1/(C_r \omega_r^2)$) from $\omega=1/\sqrt{LC}$.
Coupling strength $g$ between the qubit and measurement line 
is given by~\cite{Jeffery}:
\begin{equation}
\frac{g}{2\pi}=\frac{1}{2}\frac{C_g}{\sqrt{C_qC_r}}\sqrt{\omega_q\omega_r}
\end{equation}
When we apply $C_g=0.1$~fF, $C_q=30$~fF, $C_r=20$~fF, $\omega_q/2\pi=6$ GHz, and $\omega_r/2\pi=8$ GHz,
we have $g/2\pi\sim 0.0141$ GHz, which satisfies 
the dispersive readout $g\ll \Delta$.

The number of photons $n$ in the readout resonator is estimated as $n = P/\kappa \hbar \omega_r$, 
where $P$ denotes the input microwave power and $\kappa$ denotes the photon decay rate in the transmission line  
whose typical resonant frequency is $\omega_{cv}$.
In contrast, $P$ can be calculated using the input current $I$ and resistance $R_0$ 
given by $P=R_0I^2$:
When considering $R_0=50~\Omega$ and a typical transmission line decay rate $\kappa/2\pi=$ 1~MHz 
with $\omega_{cv}/2\pi=3$~GHz ($Q$ = 3000), 
we have
$I=\sqrt{\hbar\omega_{cv}\kappa/R_0}\approx$ 0.08 nA. 
%　=\sqrt{3.16371\times 10^{-25}J/(10ns*50Ω)]}

The ratio of the peak $\omega_p$ to the full width $\delta \omega_p$ at the half maximum is 
given by $Q_p\equiv \omega_p/\delta \omega_p$. 
Moreover, the lifetime $T_1^{\rm (m)}$ of the qubit is expressed as 
$Q_p=\omega_p T_1^{\rm(m)}$. 
Thus, this can be expressed as in Ref.~\cite{TanaAPE},
\begin{equation}
T_1^{\rm(m)} \equiv \frac{1}{\delta \omega_p}.
\end{equation} 
The output peak of the transmission line was examined using the output voltage $V({\rm out})$. 
Subsequently, $\Gamma_1$ in Eq.(\ref{fidelity}) is given by $\Gamma_1=1/T_1^{\rm(m)}$.

In conventional qubits, each qubit is controlled independently 
by switching off the interactions between qubits.
In superconducting qubit circuits, each qubit circuit is constructed separately. 
However, because all qubits are constructed on the same substrate, the parasitic capacitance cannot be neglected. 
Therefore, we consider the \(N\) qubits to be capacitively coupled.

\begin{figure}
\centering
\includegraphics[width=8.8cm]{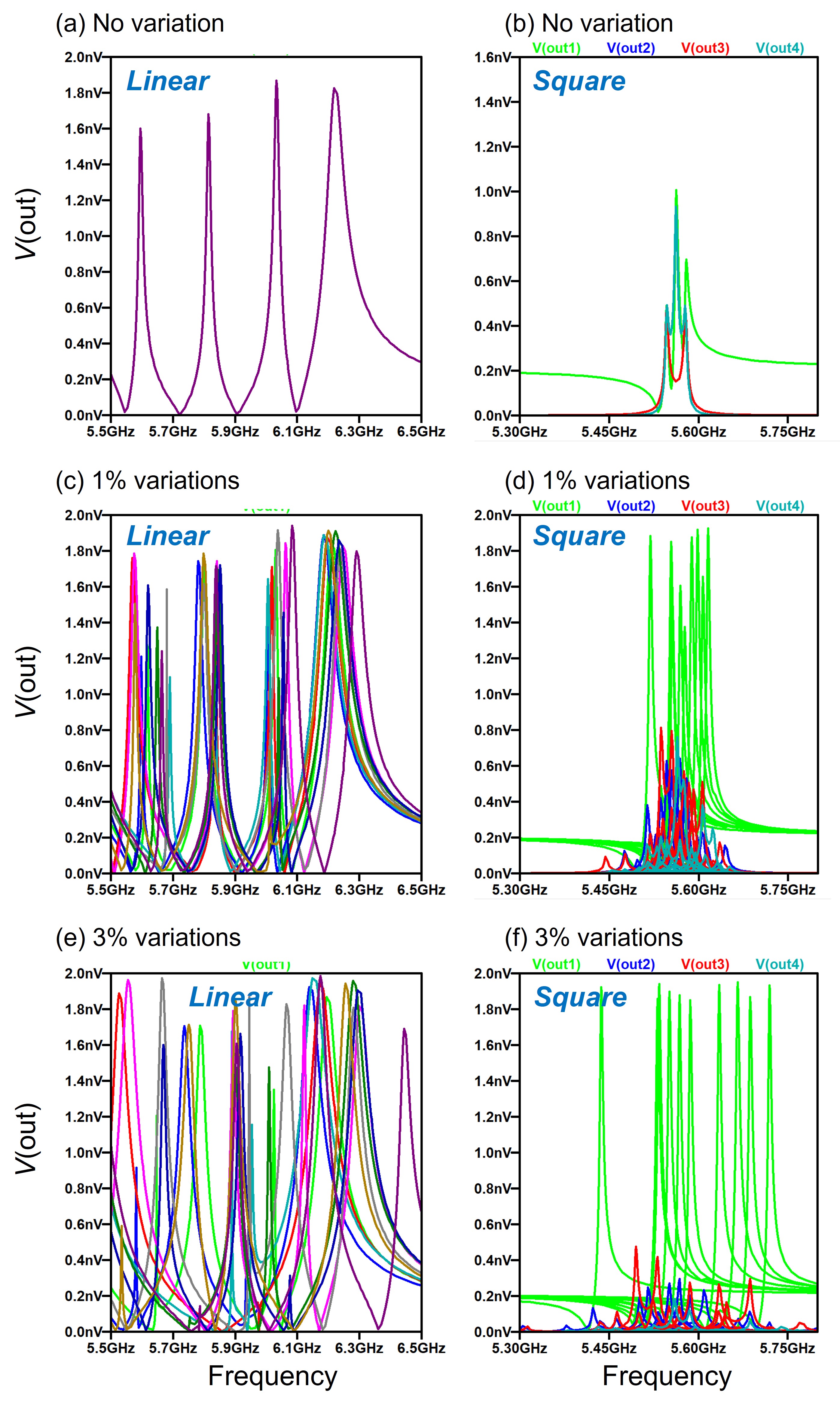}
\caption{
Simulated transmission spectra of square arrangement (a)--(c) and 
linear arrangement (d)--(f) of qubits.
Monte Carlo simulations using ten random numbers.
(a)(d) No variations (b)(e) 1\% gaussian variations are included into 
all parameters. 
(c)(f) 3\% variations are included into all parameters.
In (d)--(f), "out1"--"out4" are the outputs in Fig.2(e).
}
\label{variations}
\end{figure}

\begin{figure}
\centering
\includegraphics[width=8.5cm]{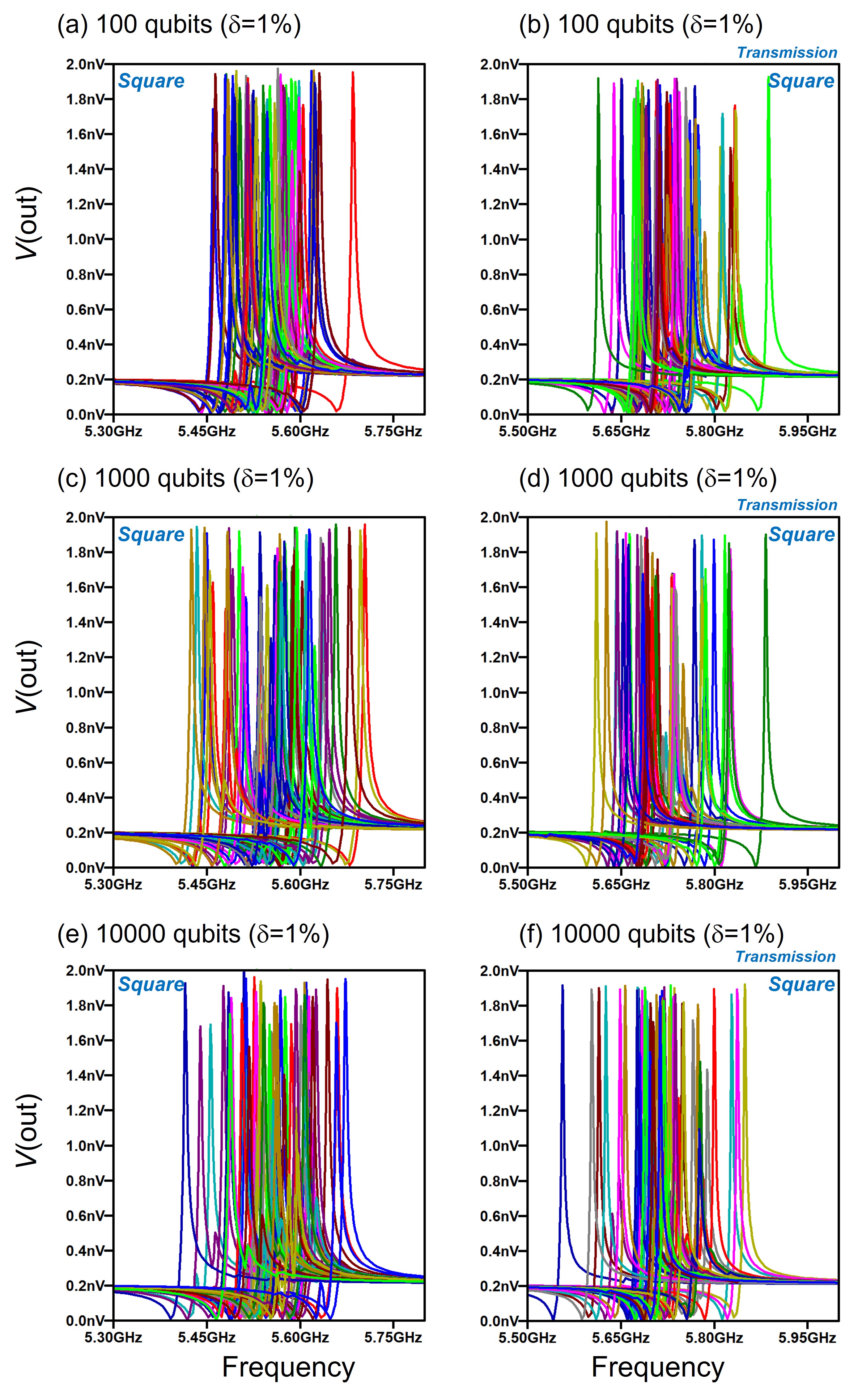}
\caption{
Examples of simulated transmission spectra of the square arrangements for several qubits. 
(a)--(c) for no transmission line. 
(d)--(f) for lossless transmission line between the qubits and the input (output).
(a)(d) for 100 qubits, (b)(e) for 1000 qubits,
and (c)(f) for 10000 qubits.
}
\label{10000qubits}
\end{figure}

\begin{figure}
\centering
\includegraphics[width=8cm]{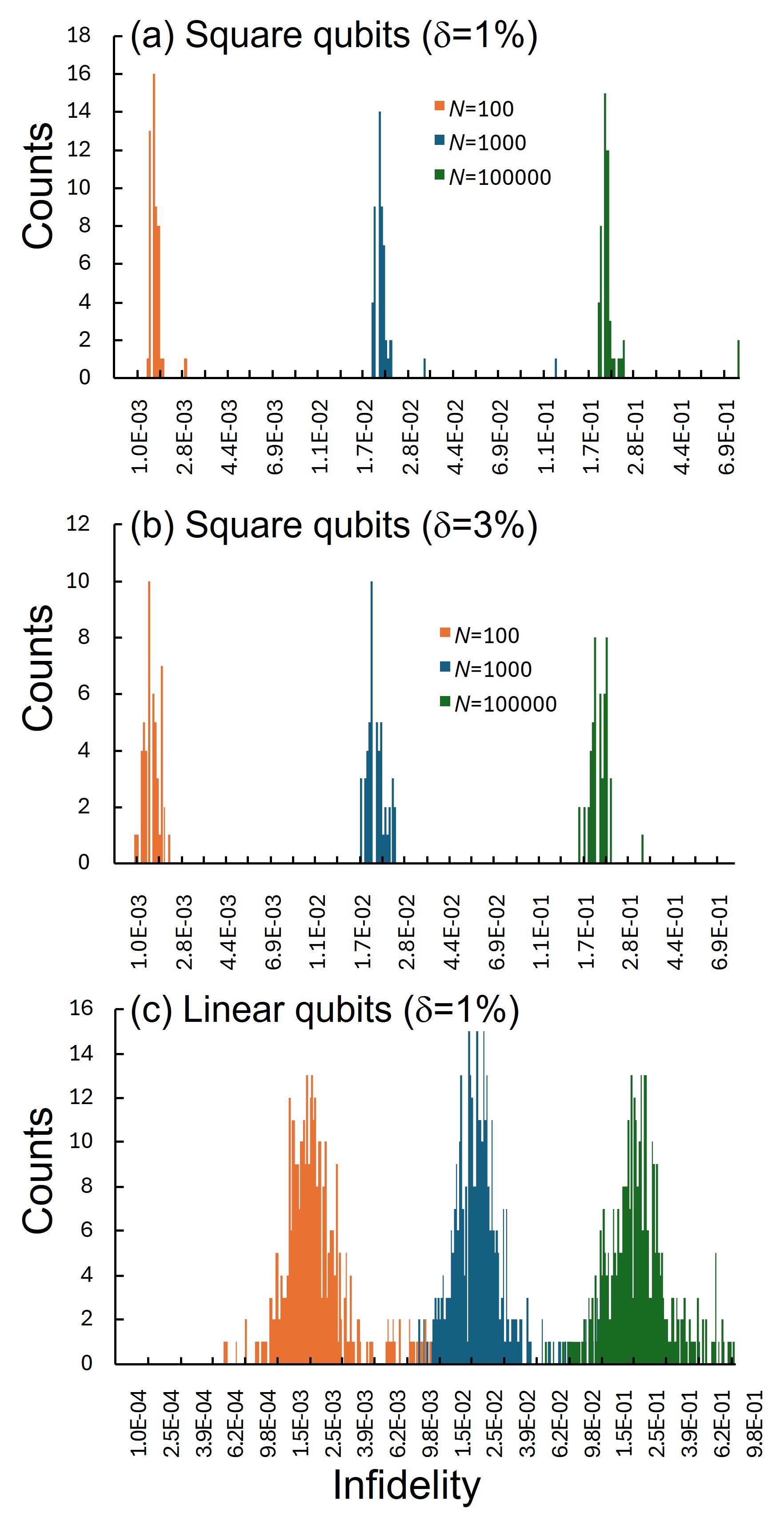}
\caption{
Infidelities calculated from Eq.(\ref{fidelity}) for 
(a) 1\% variations (b) 3\% variations of the square arrangement,
(c) 1\% variations of the linear arrangement.
$\tau_{\rm iop}=10^{-11}$ s for (a)(b) and $\tau_{\rm iop}=10^{-12}$ s for (c).
}
\label{InF}
\end{figure}

%\section{Results}
Figures~\ref{variations}--\ref{InF} show the numerical results obtained using LTspice~\cite{LTspice}.
The transmission spectra were plotted to illustrate the distribution of the resonant peaks in response to the input signals. 
Two types of peaks are observed in the AC transient analysis.
When $\omega_r > \omega_q$, the lower peaks correspond to the qubit frequencies 
and the larger peaks correspond to the resonator frequencies.
The difference between the two peaks is used for the detuning $\Delta$.
Here, we present the resonator peaks.
In Fig.~\ref{variations}, 
we observe that the transmission spectra are 
strongly affected by variations in four qubit system 
in the linear (Fig.2(c)) and square (Fig.2(d)(e)) arrangements.
All parameters ($\omega_q$, $\omega_r$,$T_1^{(q)}$, $C_{q}$, $C_{g}$, and $C_{r}$) 
include variations in central values.
For instance, the capacitances $C$ are given by $C=C_0(1+{\rm gaussian}(\delta))$.
Figures~\ref{variations}(a) and (b) 
show the output voltages for no variations for the four linear and square qubits, respectively.
In Fig.~\ref{variations}(a), the four peaks represent the four frequencies $\omega_{ri}$ of the four qubits ($i=1,..,4$).
Figure~\ref{variations}(b) shows the four outputs of the square structure, 
where only the upper left circuit had an AC power source.
Split peak structures are observed in Fig.~\ref{variations}(b) caused by the capacitive coupling (with a coupling capacitance of 0.1 fF).
These split peak structures are the result of the coupling among the circuits 
comprising bonding and antibonding states.
Figures~\ref{variations} (c)--(f) show the effects of Gaussian 
variations over ten Monte Carlo simulations.
As the variation increases from $\delta=1\%$ to $\delta=3\%$,
the peak structures vary significantly. 
In particular, for the linear structure shown in Fig~\ref{variations}(e), 
peaks for different qubits could not be distinguished. 
The distance between the two qubit frequencies is 200 MHz,
 which is 3.3\% of the qubit frequency of $\omega_{q1}/2\pi=$6 GHz.
 Thus, the difference in qubit frequencies in the linear 
 structure should be larger than the possible fabrication variations.
Precise control of the qubit parameters is required to avoid such mixture of qubit signals.

%%%%%%%%%%%%%%%%%%%%%%%%%%%%%%%%%%%%  Fig.3

%%%%%%%%%%%%%%%%%%%%%%%%%%%%%%%%%%%%  Fig.4
%%%%%%%%%%%%%%%%%%%%%%%%%%%%%%%%%%%%
Figures~\ref{10000qubits}(a)(c)(e) show the calculated outputs for 100, 1000, and 10000 qubits for the linear qubits (Fig. 1(c)), 
based on 50 repeated Monte Carlo simulations. 
In this setup, the qubit elements were capacitively coupled with a coupling capacitance of 0.1 fF.
In Figs.~\ref{10000qubits}(b)(d)(f), two lossless transmission lines are added to both the input and output. 
Consequently, the qubit section is connected to the input and output through these transmission lines, 
each introducing a delay of 10 ns. 
When comparing Figs.~\ref{10000qubits}(a)(b) with Figs.~\ref{10000qubits}(c)(d), 
we observed that increasing the number of qubits from 100 to 1000 enhanced the complexity of the peak structures. 
However, the complexities observed in Figs.~\ref{10000qubits}(c) and (d) are similar to those in Figs.~\ref{10000qubits}(e) and (f).
Thus, the effects of variations appear to be similar irrespective of the 
lossless transmission line between the qubits and the input and output.

%%%%%%%%%%%%%%%%%%%%%%%%%%%%%%%%%%%%  Fig.5
%%%%%%%%%%%%%%%%%%%%%%%%%%%%%%%%%%%%
Figure~\ref{InF} shows the calculated infidelity for the two circuit configurations. 
As the variations $\delta$ increase from Fig.~\ref{InF}(a) to (b), 
the range of the infidelity distributions was also extended. 
Although the output variations (Fig.~\ref{10000qubits}) appear similar between 1000 and 10000 qubits, 
the infidelity distributions differ because of the prefactor in Eq.(\ref{fidelity}). 
Some ambiguity regarding the parameter $\tau_{\rm op}$ is noted, 
and the values of the infidelities increase (decrease) as $\tau_{\rm op}$ increases (decreases).
Therefore, the absolute value of the infidelity should be determined experimentally.
Here, we discuss the relative values of the different configurations.
The infidelity of the linear structures (Fig.~\ref{InF}(c)) was considerably worse compared with those of the square structures (Figs.~\ref{InF}(a)(b)). 
This difference arises because, in the latter case, the four time variations have a significant impact on the output.

Finally, we briefly describe the RB process shown in Fig.1.
In the RB process, a series of single-qubit rotations and qubit--qubit operations 
was applied to the initial state~\cite{Knill,Ryan}. 
We assume $|\Psi_0\rangle=\alpha_0 |0\rangle +\beta_0 |1\rangle$.
If each single-qubit rotation and two qubit--qubit operation 
matrices include decoherence and variations in target values,
the final density matrix deviated from the pure state.  
Assume that the final density matrix resulting from the RB process is given by 
$\rho_{\rm RB} =
\left(
\begin{array}{cc}
A &  B \\
B^* & C
\end{array}
\right).
$
However, in general, the 
the time evolution of the system using the Bloch--Redfield density matrix expression is given by~\cite{Krantz} 
\begin{equation}
\rho (t) =
\left(
\begin{array}{cc}
1+(|\alpha_0|^2-1)e^{-\Gamma_1 t} &  \alpha_0 \beta_0^* e^{i\delta \omega t}e^{-\Gamma_2 t} \\
\alpha_0^* \beta_0 e^{-i\delta \omega t}e^{-\Gamma_2 t} &  |\beta_0|^2 e^{-\Gamma_1 t}
\end{array}
\right)
\end{equation}
Starting with $|\Psi_0\rangle$. 
The longitudinal and transverse relaxation rates
$\Gamma_1$ and $\Gamma_2$ represent the RB results.
We can obtain the values of $\Gamma_1$ and $\Gamma_2$ using the RB process by comparing this equation with the results of the RB calculations.
For instance, when we start with $\alpha_0=\beta_0=1/\sqrt{2}$, by comparing the elements of the 
equation $\rho(t_f)=\rho_{\rm final}$, we obtain 
$\Gamma_1 = \ln (2C)/t_f$, 
$\Gamma_2 = \ln (2|B|)/t_f$, and 
$\tanh \delta \omega t_f = {\rm Re} B/{\rm Im} B$.
We used $\Gamma_1 = \ln (2C)/t_f$ for the SPICE circuit calculations.
However, in this study, $T_1^{(q)}=\Gamma_1^{-1}=1~\mu$.

%\section{Summary}
In conclusion, designing a chip that includes multiple qubits and circuits is an extremely complicated engineering process.
This is because fabrication cannot be controlled at the atomic level and 
numerous variations cannot be avoided.
The SPICE simulator enabled us to calculate the overall performance of the circuits 
considering the margin of variation in the element devices.
Here, we provided the simplest method that 
can be used to check the effects of variations in the element devices.
Superconducting circuits, including other important components, 
such as the mixer and amplifier, can be prepared using the appropriate SPICE netlists.
Comparison with the experiments would be required in the near future to directly apply the circuit model.

%%%%%%%%%%%%%%%%%%%%%%%%%%%%%%%
\section*{Acknowledgements}
This study is partly based on the results obtained from the project JPNP16007, 
commissioned by the New Energy and Industrial Technology Development Organization (NEDO) of Japan.

%\section*{References}

%\endrefs

\end{document}